\documentclass[12pt]{article}
\usepackage{amsmath, amsthm, amssymb}
\usepackage{enumerate}
\usepackage{color}
\usepackage{graphicx}
\usepackage{float} \restylefloat{figure}
\usepackage{subfigure}
\theoremstyle{plain}
\newtheorem{thm}{Theorem}

\begin{document}

\begin{center}
{\bf {\Large Queueing Analysis of a Chagas Disease Control Campaign }}\\

\vskip0.1in
Maria T.\  Rieders \\
{\tiny rieders@wharton.upenn.edu}\\
{\tiny Operations and Information Management Department, University of Pennsylvania, Philadelphia, PA, USA }\\
\vskip0.1in

Patrick Emedom-Nnamdi\\
{\tiny patricknnamdi2@gmail.com}\\
{\tiny Department of Biology, School of Arts and Sciences, University of Pennsylvania, Philadelphia, PA, USA }\\
\vskip0.1in

Michael Z.\  Levy\\
{\tiny mzlevy@mail.med.upenn.edu} \\
{\tiny Center for Clinical Epidemiology \& Biostatistics, Department of Biostatistics \& Epidemiology}\\
{\tiny University of Pennsylvania School of Medicine, Philadelphia, PA, USA }\\
\vskip0.1in

\end{center}
\par

\begin{abstract} A critical component of preventing the spread of vector borne diseases such as Chagas disease are door-to-door campaigns by public health officials that implement insecticide application in order to eradicate the vector infestation of households. The success of such campaigns depends on adequate household participation during the active phase as well as on sufficient follow-up during the surveillance phase when newly infested houses or infested houses that had not participated in the active phase will receive treatment. Queueing models which are widely used in operations management give us a mathematical representation of the operational efforts needed to contain the spread of infestation. By modeling the queue as consisting of all infested houses in a given locality, we capture the dynamics of the insect population due to prevalence of infestation and to the additional growth of infestation by redispersion, i.e. by the spread of infestation to previously uninfested houses during the wait time for treatment. In contrast to traditional queueing models, houses waiting for treatment are not known but must be identified through a search process by public health workers. Thus, both the arrival rate of houses to the queue as well as the removal rate from the queue depend on the current level of infestation. We incorporate these dependencies through a load dependent queueing model which allows us to estimate the long run average rate of removing houses from the queue and therefore the cost  associated with a given surveillance program. The model is motivated by and applied to an ongoing Chagas disease control campaign in Arequipa, Peru.

\end{abstract}

\section{Introduction}

Control of {\it Triatoma infestans}, an insect vector of {\it Trypanosoma cruzi} which causes Chagas disease, is one of the greatest public health achievements of the past couple of decades. Residual insecticide application has greatly reduced the prevalence of the insect in areas with strong control programs. The insect however, possesses the potential to rebound from these concerted efforts. It has, on at least one opportunity, developed resistance to the commonly applied class of insecticides (the pyrethroids). The insect has also encroached into urban habitats.The size and complexity of cities possess numerous challenges to vector control. Insects can move easily between habitats in densely populated cities. Participation in control efforts is always partial, and non participating households can serve as foci for the return and re-dispersion of vectors following control efforts. The city of Arequipa, Peru, with a population of nearly one million is probably the largest city battling T.\ infestans. Insecticide application in the city began in 2003; to date over 80,000 households of the city have been treated.  All but two of the affected districts of the city have passed from the {\it attack phase} of the campaign - during which insecticide is applied twice at a six month interval - into the {\it surveillance phase}. The surveillance phase relies on a combination of community reporting of returning insects and active search by trained entomological surveyors. 
\par

In a traditional queueing model with one or more servers, customers arrive at a service facility according to some stochastic arrival process. If all servers are busy at time of arrival, the customer will join the end of the queue, waiting for his/her turn in service. Service may be given on a first come, first served basis, a last come, first served basis, by randomly selecting customers from the queue, or by some other selection policy. Classic queueing models include the $M/G/1$ queue where jobs arrive according to a Poisson process, are being served by a single server with the sequence of service times being independent and identically distributed (iid) random variables. When service times are required to be iid exponential, such a system is called an $M/M/1$ queue. In these standard systems, we assume that jobs joining the queue will not renege, and that all underlying distributions remain the same throughout time and are independent of each other. These basic systems are well understood in terms of performance measures such as waiting time of a job, throughput analysis or queue length distribution. For results on basic queueing systems, the reader may refer to Kleinrock \cite{Kleinrock1975} or Cooper \cite{Cooper1981}.
\par
In the context of our public health campaign, customers requiring service are houses that have been infested with  Triatoma infestans, the disease carrying insect. Treatment of such a customer (house) consists of identifying an infested house and having it sprayed by exterminators with a pesticide that will eradicate the insect population on the premises. Thus, we imagine infested houses to be part of an imaginary queue where they will remain until they have been identified as hosting the Triatoma infestans and been treated by insecticides. We may also think of the houses in queue as {\it invisible} houses, recognizing the challenge for the public health workers to make them visible and then remove them from the queue.
\par
Our queueing model tries to capture the resources needed to deal with infested houses during the surveillance phase. For describing the arrivals to our imaginary queue, we note that houses currently in the queue may have become infested - and thus joined the queue - in two different ways: The insect vector may have been brought into a household from outside the neighborhood. Furthermore, any of the infested houses in the queue is capable of infesting other houses by spreading the vector into previously uninfested houses; the rate of this secondary infestation depends on how long a house has been infested. Thus, the arrival rate of houses is composed of an external rate plus a redispersion rate that depends on how long houses remain in the queue. In order to remove a house from the queue, public health officials need to identify that a particular house is infested. This will be accomplished by a multi-armed bandit algorithm that searches for infested houses based on historical data and continuously updated GIS maps; see \cite{Gutfraind2015}. The total service time, i.e. the time to remove a house from the queue, consists of the time until an infested house has been identified plus the time required for actual treatment with pesticides.
\par
The connection between queueing models and epidemic models has been addressed in the literature by Trapman and Bootsma \cite{Trapman2009} and Hern\'andez-Suarez et al.\ \cite{Hernandez-Suarez2010}.  \cite{Trapman2009} uses results about a classic queueing system (the M/G/1 queue with processor sharing) to estimate the distribution of the number of infectives at the moment of first detection in a stochastic epidemic model of SIR type, where SIR stands for Susceptible-Infectious-Removed/Re\-covered individuals. \cite{Hernandez-Suarez2010} study SIS (Susceptible-Infected-Susceptible) and SEIS (Susceptible-Exposed-Infected-Susceptible) epidemic models and find the distribution of the disease while it is in the endemic state by using the classic M/G/N queueing system. Kaplan et al.\ \cite{Kaplan2003} built a trace-vaccination queue to model operational requirements for dealing with a smallpox epidemic. Lee et al. \cite{LeeInterfaces2015} incorporate a disease propagation model with a vaccine queueing model in order to develop prioritization rules for the use of limited vaccines available for containing a pandemic. Both \cite{Kaplan2003} and \cite{LeeInterfaces2015} use systems of ordinary differential equations for analyzing the trajectory of a pandemic outbreak over a finite time horizon. In contrast, our work is concerned with analyzing the continuing surveillance of infestation over a long period of time.  In particular, our model will enable public health officials to estimate long run operational efforts and therefore the cost of maintaining a public health campaign which is charged with containing a prevalent infestation.

\par
In Section 2, we will present the queueing model in detail, including how to estimate the redispersion rate as a function of the cumulative prevalence of infestation in the area under consideration. We will introduce the concept of total workload of a queueing system and formulate both the arrival rate as well as the removal rate of houses as functions of the workload. Section 3 presents the long run analysis of such a load dependent queueing system, including an algorithm for calculating the operational throughput and the average cost per time period for treating infested houses during the surveillance phase. Several special cases that allow for closed form or simplified algorithmic calculation are considered. Computational results are based on field work in Arequipa, Peru and illustrate the implementation and data requirements for our analysis. In Section 4, we summarize our contributions and describe future research directions.
\par
\section{The Queueing Model}
\begin{figure}[h!] \caption{Queueing Model} \centering \includegraphics[width=1
\textwidth]{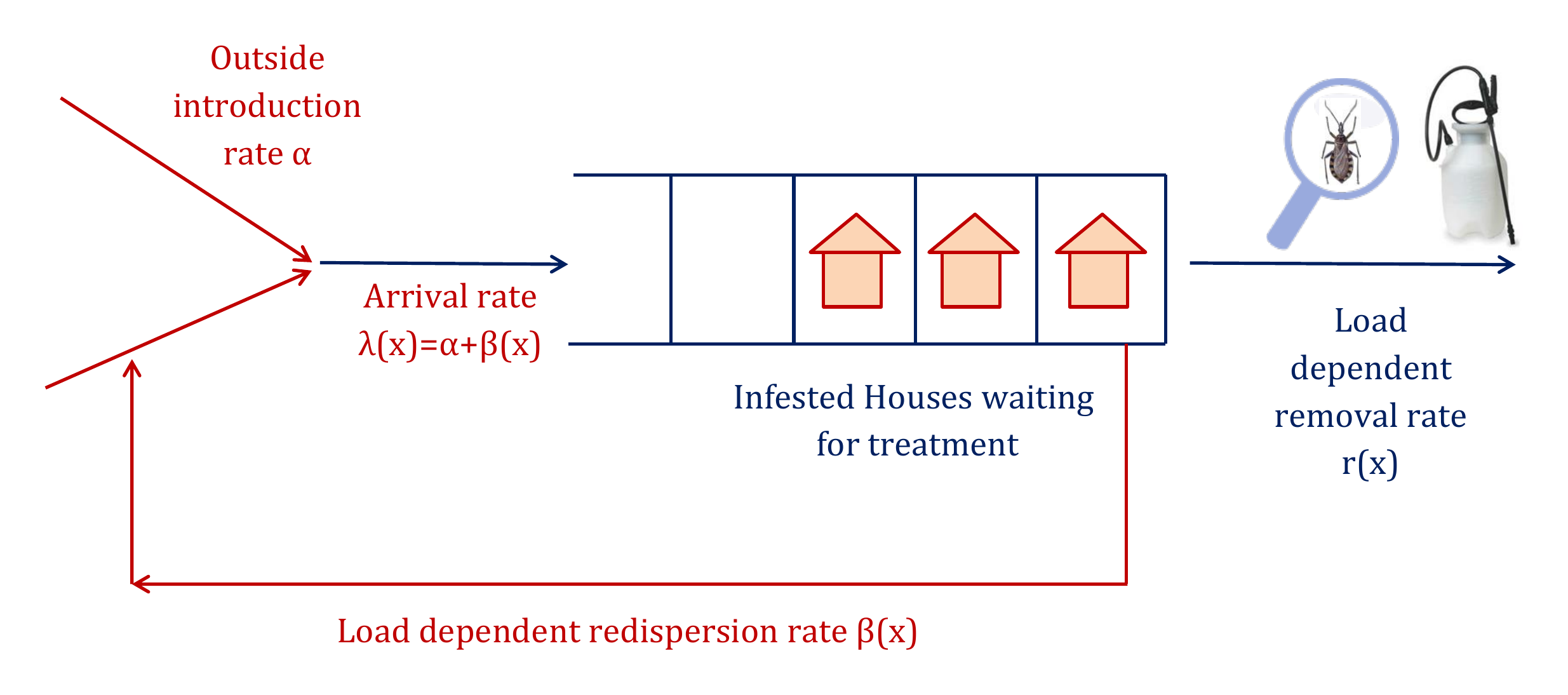} \label{fig:Queue}\end{figure}

In the following, we describe the basic mechanics and assumptions of our queueing model. Note that the classic operations management literature use the words customers or jobs for entities flowing through a queueing system. In our context, customers are infested houses; the arrival of a customer corresponds to a house becoming infested and a service time completion is equivalent to identification and treatment of an infested house.  In other words, our model envisions all currently infested houses as having joined an imaginary queue at the time of their initial infestation. A house will remain in queue until its infestation has been discovered and verified by a public health inspector and the house has been subsequently treated with an insecticide that will eliminate infestation and prevent further spread of infestation form this house. A schematic representation of this model is given in Figure  \ref{fig:Queue}.  The next subsection will state the technical assumptions that we impose on the system.
\par

\begin{figure}
\centering
\begin{subfigure}{}
  \centering
  \includegraphics[width=2.5in]{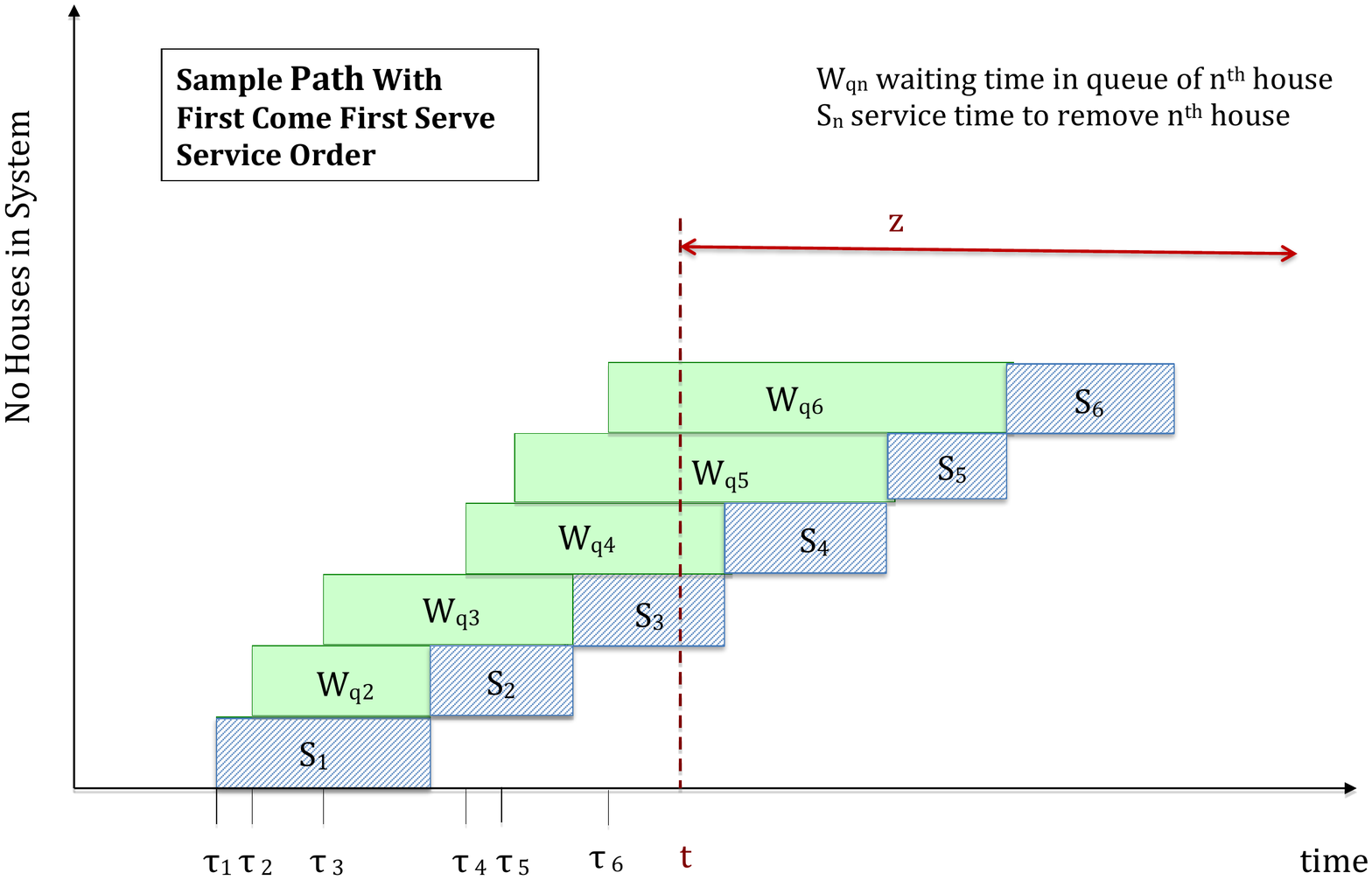}
\end{subfigure}%
\begin{subfigure}{}
  \centering
  \includegraphics[width=2.5in]{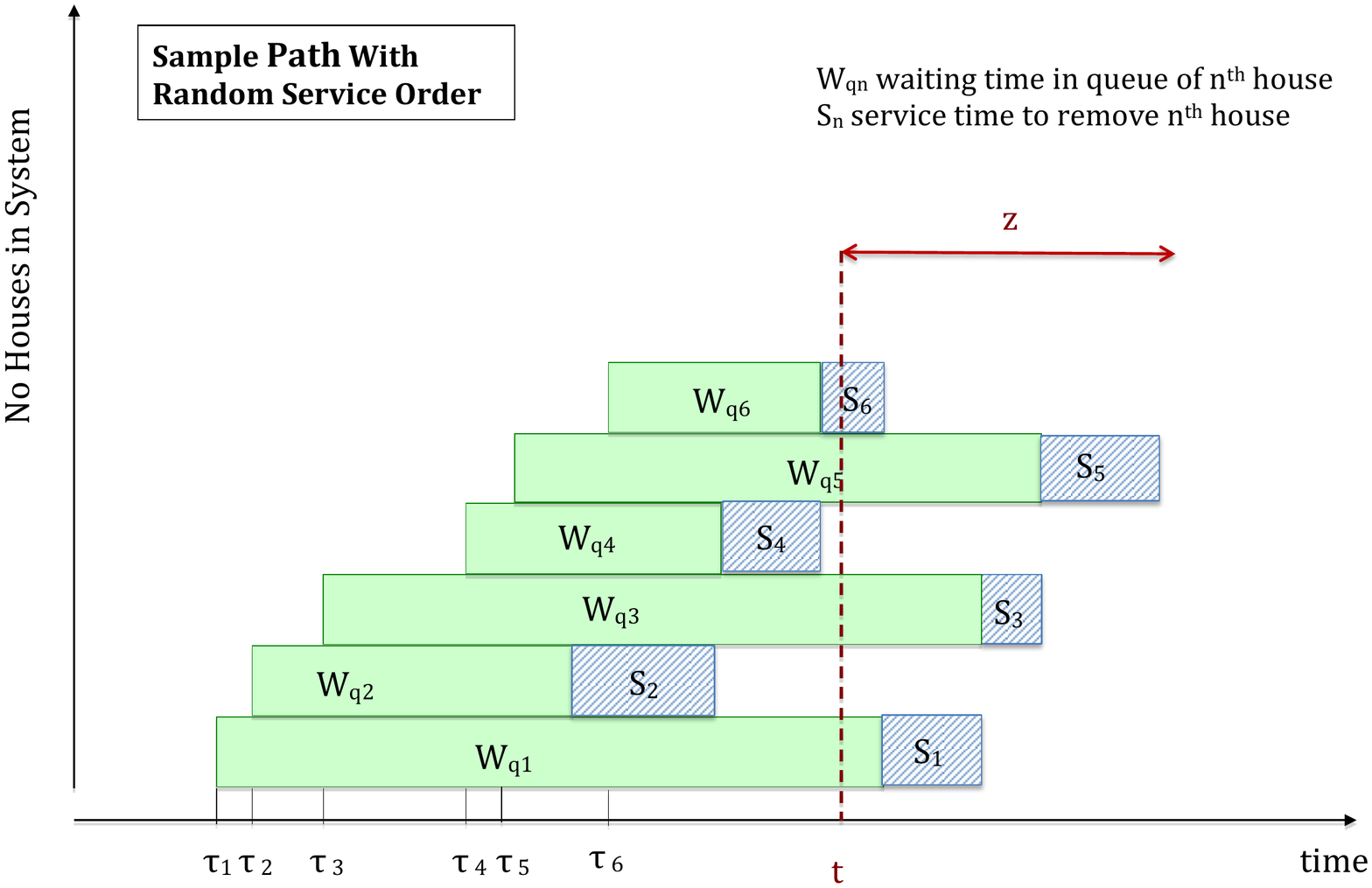}
\end{subfigure}
\caption{Sample Paths for Different Service Disciplines}
\label{fig:test}
\end{figure}

Figure \ref{fig:test} shows a typical sample path of a queueing system where jobs arrive at times $\tau_1,\tau_2,\ldots$ to the system. The $i^{\rm th}$ arriving job may have to wait an amount of time $W_i$ before receiving a service time $S_i$. Upon completion of the service time, job $i$ will then leave the system.

\subsection{Assumptions of the Queueing Model}
\par
Infestation of households with the {\it Triatoma infestans} typically happens either by the introduction of the insect vector from the outside or from insects spreading from infested properties to noninfested households. We assume that arrivals from the outside occur according to a Poisson process with a flat rate $\alpha>0$. In addition to imported infestations from outside our geographic area, we note that the insects living in an infested house will spread their colonies to neighboring properties, thus adding previously uninfested houses to the queue. We call the rate of this spread of infestation the {\it redispersion rate} $\beta$ and note that $\beta$ depends on how many households are currently in the queue and how long each of them has been waiting for treatment. Hence, the total arrival rate is given by
\begin{equation}\lambda(x)= \alpha+\beta(x)\end{equation}         
where $x$ stands for the current occupancy state of the system. In a very detailed model, $x$ would include information on the number of infested houses as well as the duration of infestation for each one of these houses since the geographic spread of insects is a function of time. We will propose to use an aggregate measure instead, namely the virtual load; see Subsection 2.1 below.
\par
Houses are removed from the queue through a process consisting of search, identification, and treatment. When only a few houses are infested, i.e., when the level of infestation is low, the search to identify infested houses will take longer than during times of more severe infestation. We therefore assume that the {\it removal rate} $r(x)$ at which houses are identified and treated from the queue is also dependent on the current state $x$ of the system (analogous to the redispersion rate function $β(x)$). Note that the exogenous arrival process with rate $\alpha$ is independent of the redispersion arrival process with rate $\beta(x)$. Thus, houses join the queue at a composite rate of
\begin{equation}
\lambda=\alpha+\beta.
\end{equation}
\par

\par
\subsection{The Redispersion Rate Function}
\begin{figure}[h!] \caption{Infectivity As A Function of Time} \centering \includegraphics[width=0.5
\textwidth]{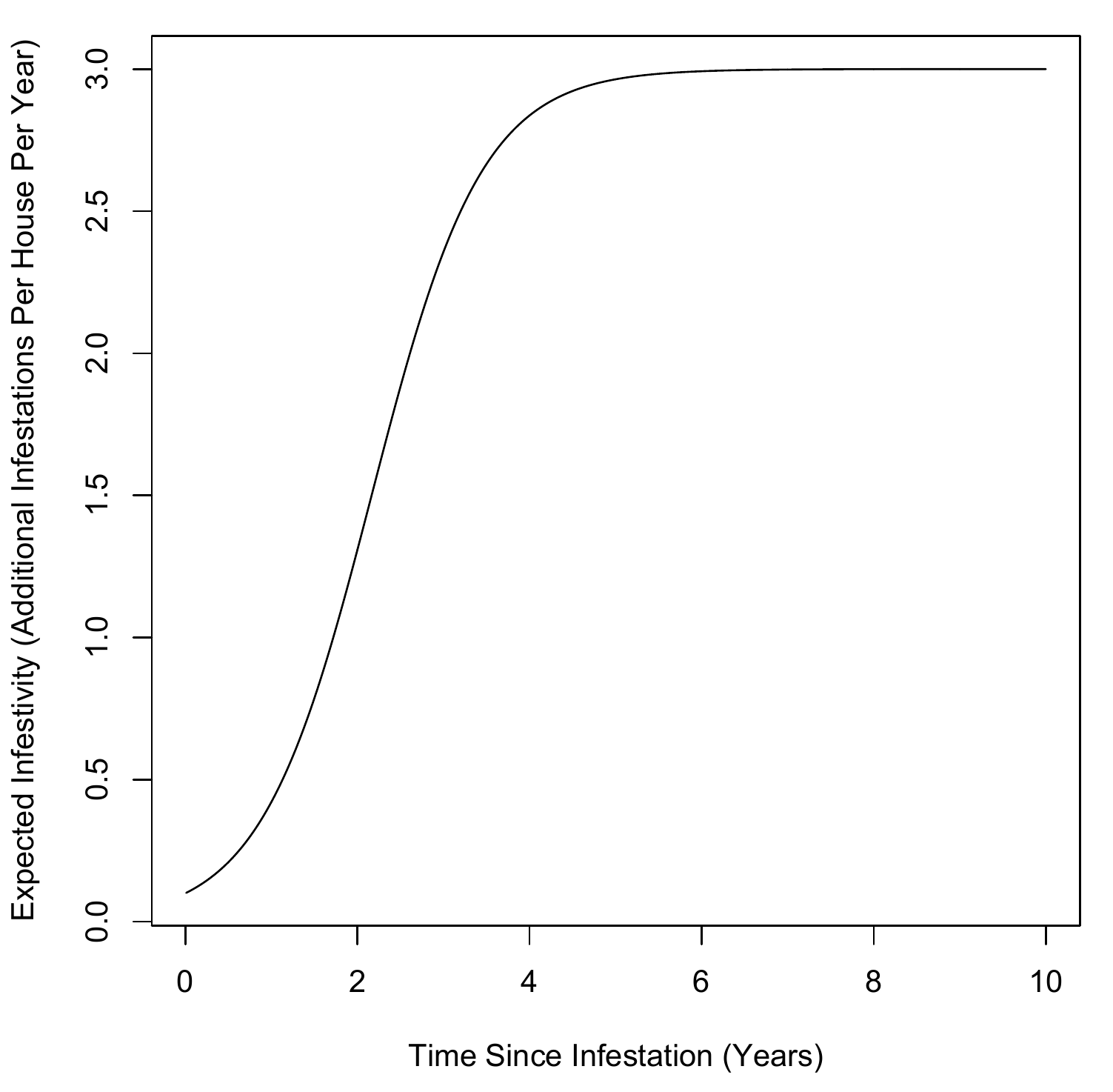} \label{fig:Infectivity}\end{figure}

We estimate the redispersion rate function based on empirical work on the dynamics of the insect population growth as a function of time. Rabinovich \cite{Rabinovich1972} performs a statistical evaluation of the population dynamics of {\it Triatoma infestans}, the insect responsible for carrying the Chagas disease parasite. Based on Rabinovich's work, lab experiments, and the empirical data collected in Arequipa, Barbou et al.\ \cite{Barbu2013}, \cite{Barbu2014}, and \cite{Barbu2015} conceptualized the dispersal of the insect through the city as a series of "hops", "skips" and "jumps". Using simulated data and longitudinal data from sequential surveys of numerous localities across Arequipa they were able to make inference on how the bug moves through the city. They found that, on average, it takes 2 years for an infested household to successfully infest an additional household. This slower than expected rate is good news as it means that public health inspectors have some time to detect and eliminate new infestations, and partially explains why such an imperfect control campaign has been extremely successful. Based on their work, we can estimate the additional number of houses that will be infested by one untreated house to follow a logistic curve as a function of time since infestation; see Figure \ref{fig:Infectivity}.

Let us denote the infectivity function of Figure \ref{fig:Infectivity} by $\gamma(t)$. Note that our queueing model requires us to estimate the redispersion function $\beta$ as a function of the overall state of the system.  For instance, if at time $t$, there are $n$ houses in queue, with respective arrival epochs $\tau_1,\tau_2,\ldots, \tau_n$, then each of these houses will reinfest other houses at rate $\gamma(t-\tau_i)$, $i=1,\ldots,n$.  Assuming no overlap in the additional infestations, this would result in a cumulative redispersion rate 
\begin{equation} \label{eq:betaDetail}
\beta(x)\approx \sum_{i=1}^n \gamma(t-t_i).
\end{equation}
Note, however, that the analytical model does not keep track of the actual arrival times of the houses currently in the queue nor the current queu length. In the following, we therefore descibe a way to approximate $\beta$ as a function of the {\it total work} in the system. In queueing systems, the total work, also called {\it workload} $V(t)$ refers to the sum of all service times of the jobs currently waiting in queue plus the remaining service time of jobs currently in service. In a single server queueing system, the total work is also referred to as the {\it virtual wait}. We propose to use the workload as a surrogate for the state $x$ of the system. In particular, we express the redispersion rate β as a function of the current workload. That is, we assume that we know the virtual load  $x$, but we do not know the current number $n$ of infested houses. From Little's law (see e.g., \cite{Wolff1989}) we know that the average number of jobs in a queue equals the average arrival rate times the average time spent in queue. Hence, we approximate the number of houses in the system $n$ by
\[n\approx \lambda x.\]
Since the time $\tau-t_i$ refers to the age of the infestation of house $i$ and we consider the queueing system in the long run, we will treat it like the long run expected age of a renewal process with renewal intervals of size $x$, that is $\frac{z^2}{2z}$, obtaining the following approximation for the redispersion rate $\beta$
\begin{equation} 
\beta(z)\approx\lambda z \gamma(\frac{z}{2}).\end{equation}
We note that the factor $\lambda$ is the long run average arrival rate of houses to the queue, a quantity that we do not know a priory as it depends on the stationary flow through the system through the redispersion feedback look. We know that the approximation (\ref{eq:betaDetail}) will tend to overestimate the new infestations since currently infested houses may be clustered together, thus reducing the number of most likely candidates for reinfestation that are surrounding an existing insect colony. On the other hand, the rate $\lambda$ itself is bounded from below by $\alpha$. Thus, as a first approximation of $\beta$, we propose to use
\begin{equation} \label{eq:betaApprox}
\beta(x)\approx \alpha z \gamma(z/2)
\end{equation}
and for the overall arrival rate
\begin{equation} \label{eq:lambdaApprox}
\lambda(x)\approx\alpha+\alpha z \gamma(z/2).
\end{equation}
The suitability of these approximations is currently being investiated through simulation experiments.

\subsection{The Treatment Rate} 
Successful treatment of infested households and therefore the removal of that household from our queue does require identification of infested house by trained inspectors. The public health campaign in Arequipa has collected data over years that allows the ranking of houses based on risk factors for reinfestation. Billig \cite{Billig2017} has developed a Baysian algorithm that updates these risk maps continuously using new data from field workers as they embark on daily searches. Thus, field workers have access to ranked lists of most likely infested houses and may choose to visit homes based on the given rankings. If the search is successful in identifying an infested house, treatment is scheduled for that property and the house thus removed from the queue. Currently, these search algorithms are being simplified and made available on mobile phones through an app. Regardless of which particular algorithm is employed, it is clear that, in general, a successful search for and treatment of an infested house will depend on the prevalence of infestation. A geographic area which contains only a few infested households is likely to require a longer search for vector colonies while a heavily infested area will typically require less effort in locating infestations. Therefore, we assume in our queueing model that the removal rate $r(x)$ will depend on the level of infestation. Just as with the redispersion rate, we take the current workload $x$ as a proxy for the level of infestation. Concrete forms for the function $r(x)$ will have to be determined based on the specific search algorithm employed. In our analysis below, we focus on a few basic functional forms that give us some insight into the long run effort required for containing the spread of infestation.

\subsection{A Load Dependent Queueing Model}
Queues with workload dependent removal rates have originally been modeled like the storage of water in dams or reservoirs; see for instance Asmussen \cite{Asmu2000}. Water flows into a reservoir from one or several water sources and is released at a rate $r(x)$ depending on the present content of water $x$ in the reservoir. Hence, if $r(x)=1$, the workload decreases at rate $-1$ as long as work is in the system. If the workload at time $t_0$ is $x$ and the next arrival is at time $t_1>t_0$, the workload process during the interval $(t_0,t_1)$ can be characterized as $V(t_0+t)=x-\int_{t_0}^{t_0+t}r(s)ds$. 
\par 
In the following, we will use an $M/G/1$ type queueing model with workload dependent arrival and service rates as analyzed by Bekker \cite{Bekker2004}. When the current workload of the system is $x$, then customers arrive according to a Poisson process with rate $\lambda(x)$; i.e., the time until the next arrival $A$ after current time $t_0$ is distributed according to $P[A>t]=e^{-\int_0^t \lambda(V(s)ds}$. The function $\lambda(\cdot)$ is assumed to be nonnegative and left continuous. Each arrival to the system adds a service time $S$ to the existing workload. We assume that the sequence of service times $S_1$, $S_2$, $S_3,\ \ldots$ are independent and identically distributed (iid) random variables with cumulative distribution function $F(\cdot)$ and mean $E[S]$. While service times are iid, the rate $r(x)$ at which service is delivered is assumed to be dependent on the current workload in the system. That is, between arrivals we have $\frac{dV(t)}{dt}=-r(V(t))$. We assume that $r(0)=0$ and that $r(\cdot)$ is strictly positive and left-continuous. We may think of the sequence $\{S_1$, $S_2$, $S_3,\ \ldots\}$ as nominal service times that would be required to serve each customer (house) under some normal condition. The effect of longer search times due to sparser or less severe infestation is being modeled by the service rate being slower than for a situation when infestation is less sparse or more severe. For this reason we assume that the function $r(x)$ is increasing in $x$.

\section{Analysis of the Load Dependent Queueing Model}
Our analysis of the load dependent queueing system presented in the previous section draws heavily on the results derived in Bekker \cite{Bekker2004} and references therein. We first present these analytical results in a general setting and then empirical findings in the context our Chagas disease public health campaign.
\subsection{General Analytical Results}
Since our interest lies in the stationary analysis of the queueing system, we start with stating conditions on arrival and removal rate functions that guarantee stability. The results on stationary workload are based on level crossing arguments and can be found in \cite{Bekker2004}.
\begin{thm} The queueing system is stationary if 
\begin{equation}
\limsup_{x\to\infty} E[S] \frac{\lambda(x)}{r(x)}<1.
\end{equation}
\end{thm}
We now assume stationarity and consider the steady-state random variables $V$ denoting the workload and $W$ denoting the workload immediately before an arrival epoch. Let $v(\cdot)$ and $w(\cdot)$ be the probability density function of $V$ and $W$, respectively, and $V(\cdot)$ and $W(\cdot)$ their cumulative distribution functions.

\begin{thm}\label{WorkloadDensity}
The workload density $v(\cdot)$ exists and satisfies the equation 
\begin{equation}
r(x)v(x)=\lambda(0)V(0)(1-F(x))+\int_{y=0^+}^x (1-F(x-y))\lambda(y)v(y)dy,\quad    x>0.
\end{equation}
\end{thm}            

Note that knowledge of the workload density allows us to calculate the long run average arrival rate $\bar{\lambda}$ to the system, i.e., the average number of houses infested and treated per year.  The law of total probability gives us the following result:
\begin{equation}
\bar{\lambda}=\int_{0^+}^\infty \lambda(x) v(x) dx +\lambda(0)V(0).
\end{equation}
For a few select cases, the expressions in \ref{WorkloadDensity} can be simplified as follows.
\par
\begin{itemize}
\item[(i)] Consider a load dependent $M/G/1$ Queue with arrival rate proportional to service rate, i.e.,  $\lambda(x)=Cr(x)$. In this case, $r(x)v(x)$ equals the virtual load $v(x)$ of an M/G/1 queueing system with arrival rate $C$ and service speed 1. 
\item[(ii)] If we assume a load dependent $M/M/1$ queue with general arrival rate and service rate function, i.e., if both interarrival times and service times are exponential, albeit with rates that depend on the current workload $x$, then we can solve the differential equation in Theorem \ref{WorkloadDensity} and obtain the density function $v(x)$ as follows:
\begin{equation}
v(x)=\frac{\lambda(0)V(0)}{r(x)} {\rm exp} \left\{\int_0^x \left( \frac{\lambda(y)}{r(y)}-\mu \right)dy\right\}.
\end{equation}
\item[(iii)]
A further simplification of the system in part (ii) to an $M/M/1$ queue with load dependent arrival rate function $\lambda(x)$ and constant service rate $\mu$ results in the queueing model studied by Brill in \cite{Brill1988}. Brill's paper offers a simple computational algorithm for calculating the average throughput $\bar\lambda$ based on level crossing arguments.
\end{itemize}

\subsection{Computational Results}
We currently are implementing a simulation study to test the appropriateness of the redispersion function as estimated by Equation (\ref{eq:betaDetail}). Subsequently, we will conduct a suite of computational experiments, demonstrating the power of investigating the yearly cost of the public health campaign. In particular, we propose to investigate the following questions.
\begin{itemize}
\item How many houses will require treatment per year when the campaign applies a certain search strategy?
\item Could additional investment into search procedures result in lower annual cost for the campaign due to fewer redispersions?
\item What is the minimum effort required (in terms of search success rates) in order to contain the spread of infestation at a manageable level?
\end{itemize}

\section{Concluding Remarks}
Operational insights gained by applying queueing theory is a novel approach for studying the effectiveness and cost of a public health campaign that is focused on the long term surveillance of a geographic area. Through the rich data collection that the team in Arequipa, Peru has acquired over the years, we have gained fairly sophisticated understanding of the dynamics of the spread of the insect vector {\it Triatoma infestans}. This work is a first attempt to combine these dynamics with the operational effects of a public health team charged with searching for and eradicating residual infestations. In future work on this project, we plan to focus on better models for the removal rate $r(x)$ as we gain more insights into the actual performance of various search strategies in the field.

\newpage
\noindent{\bf Acknowledgement} 
\par\noindent The authors acknowldege programming assistance by Karthik Sethuraman (former student at the Department of Bioengineering, School of Engineering and Applied Science, University of Pennsylvania, Philadelphia, PA, USA) for initial work on this project. We also thank the following organizations for their part in organizing and conducting the Chagas Disease control campaign in Arequipa: Ministerio de Salud del Perú (MINSA), the Dirección General de Salud de las Personas (DGSP), the Estrategia Sanitaria Nacional de Prevención y Control de Enfermedades Metaxénicas y Otras Transmitidas por Vectores (ESNPCEMOTVS), the Dirección General de Salud Ambiental (DIGESA), the Gobierno Regional de Arequipa, the Gerencia Regional de Salud de Arequipa (GRSA), the Pan American Health Organization (PAHO/OPS) and the Canadian International Development Agency (CIDA). This work was supported by National Institutes of Health grants NIH-NIAID R01AI101229 and R01HD075869.
\end{document}